# Attenuation of the NMR signal in a field gradient due to stochastic dynamics with memory

Vladimír Lisý [a, b, *], Jana Tóthová [a]

[a] *Department of Physics, Technical University of Košice,*
*Park Komenského 2, 042 00 Košice, Slovakia*
[b] *Laboratory of Radiation Biology, Joint Institute of Nuclear Research,*
*141 980 Dubna, Moscow Region, Russia*

**Abstract**

The attenuation function $S(t)$ for an ensemble of spins in a magnetic-field gradient is calculated by accumulation of the phase shifts in the rotating frame resulting from the displacements of spin-bearing particles. The found $S(t)$, expressed through the particle mean square displacement, is applicable for any kind of stationary stochastic motion of spins, including their non-markovian dynamics with memory. The known expressions valid for normal and anomalous diffusion are obtained as special cases in the long time approximation. The method is also applicable to the NMR pulse sequences based on the refocusing principle. This is demonstrated by describing the Hahn spin echo experiment. The attenuation of the NMR signal is also evaluated providing that the random motion of particle is modeled by the generalized Langevin equation with the memory kernel exponentially decaying in time.

## 1. Introduction

Nuclear magnetic resonance (NMR) has proven to be an effective method of studying molecular self-diffusion and diffusion in various materials and has a wide range of applications ranging from characterization of solutions to inferring microstructural features in biological tissues [1–13]. The effect of diffusion on the NMR signal was incorporated by Torrey into the Bloch equations for the spin magnetization of an excited sample placed in a magnetic field [14]. By solving these equations the NMR signal is obtained as a product of the time evolution of the magnetization without the influence of diffusion and the diffusion suppression function $S(t)$. The function $S(t)$ can also be evaluated through the time-dependent resonance frequency offset in the frame rotating with the resonance frequency [1–4, 10, 15 –23]. Evidently, the results based on the first approach do not go beyond the Einstein–Fick theory of diffusion. The second approach needs more attention. Most calculations of $S(t)$ within this approach are valid within the long-time approximation. This means that only experiments on spin-bearing particles undergoing diffusion in liquids or gases, when their mean square displacement (MSD) is recorded at times $t$ much larger than the characteristic frictional time of the particles [24], can be interpreted within the current theories. For shorter times, beyond the diffusion regime of the particle motion, these theories are inapplicable. To our knowledge, there are only a few exceptions, when $S(t)$ is successfully evaluated for all times [25–27]. In these works, the formulas for the attenuation of the NMR signal in a field gradient have been found in the frame of the standard Langevin theory of the Brownian motion [28]. This theory, however, has strong limitations. In fact, it is applicable only in the long-time limit when, however, it is not necessary since then the Einstein theory is valid. For shorter times it works only for Brownian particles with the mass much exceeding the mass of the surrounding particles (such as dust particles in a gas) and is inappropriate for Brownian particles in a liquid [29–31]. With the aim to interpret NMR experiments at all times, new results based on a more general theory of the Brownian motion are thus highly desirable. In this paper, we show that the attempts [25–27, 20] to describe the NMR experiments by using the generalized Langevin equation (GLE) with a memory integral [32, 33] must be corrected.

In the next section, a discussion on the limitation of current calculations of $S(t)$ is given. Then we present a different way of finding $S(t)$ for an ensemble of stochastically moving spins in a magnetic-field gradient. The attenuation function will be calculated through the accumulation of the phase shifts in the rotating frame due to the particle displacements. Instead of the use of the particle positional autocorrelation function (PAF) as in Ref. [20], which is ill defined for unbounded random motion, $S(t)$ is expressed through the MSD. The obtained new formulas are independent on a model used for the description of the particle stochastic dynamics. It is only assumed that the studied random processes are stationary in the sense that the autocorrelation function of a fluctuating dynamical variable $x$ at times $t$ and $t'$ depends only on the time difference, $t - t'$: $\langle x(t)x(t')\rangle = \langle x(t-t')x(0)\rangle$ [10]. With this assumption the results are applicable to the Brownian motion with memory and at long times to normal and anomalous diffusion.

An experiment is described when the nuclear induction signal is read-out in the presence of a constant field gradient. Then the used method is developed to the NMR pulse sequences based on the refocusing principle. Namely, the Hahn spin echo experiments with steady and pulsed field gradients are described. Finally, we obtain the NMR induction signal assuming that the particle displacements can be described by the GLE, used in the literature to model the Brownian dynamics in viscoelastic (Maxwell) fluids [34, 35].

## 2. Limitations of the current calculation of the diffusion suppression function

Let us consider an experiment, in which the nuclear induction signal is observed in the presence of a steady magnetic-field gradient [20–23]. A liquid or gaseous system is placed in a strong magnetic field along the axis $x$ and after the 90° rf pulse the magnetization of an ensemble of spins is modulated by the field gradient. The influence of diffusion on the total magnetization determining the observed NMR signal is commonly expressed as [1–3, 10, 15–23]

$$S(t) = \langle \exp[i\phi(t)] \rangle = \left\langle \exp\left[i\int_0^t \omega(\tau)d\tau\right] \right\rangle, \qquad (1)$$

where $\omega(t)$ is the time-dependent resonance frequency offset in the rotating frame and the brackets $\langle ... \rangle$ mean the expectation value of the phase factor as a function of time [36]. In the literature [20], the relation $\omega(t) = \gamma_n g x(t)$ is used, where $g$ is the applied gradient strength, $\gamma_n$ is the nuclear gyromagnetic ratio, and $x(t)$ is the position of the spin after time $t$. It follows from Eq. (1) that

$$S(t) = \exp\left[-\frac{1}{2}\langle \phi^2(t) \rangle\right]. \qquad (2)$$

The variable $\phi(t)$ varies randomly over the ensemble of spins. A sufficient assumption in obtaining Eq. (2) is the Gaussian distribution of $\phi(t)$, which is, according to the central limit theorem, a good assumption when dealing with very many randomly varying quantities [10]. For a non-Gaussian distribution, Eq. (2) holds approximately for a small $\phi$. A detailed derivation of this equation is given in [23]. Then the authors use the relation $\langle x(t)x(t+T) \rangle = D(2t + T - |T|)$, which, according to [22, 23], is obtained from the Einstein-Smoluchowski equation for 1D diffusion, $\langle x^2 \rangle = 2Dt$. Here, $D = k_B T / \gamma$ is the diffusion coefficient determined by the temperature $T$ and the friction coefficient $\gamma$. From the above relation for the PAF one finds at $t = 0$ and $T > 0$ that $\langle x(0)x(T) \rangle = 0$. For 1D diffusion, an unambiguous quantity is the MSD, for which $X(t) = \langle [x(t) - x(0)]^2 \rangle = 2Dt$ holds within the Einstein theory. Although in [22, 23] the correct result has been presented for $S(t)$ (see Sec. 5), the used approach is applicable only for long times (much larger than the relaxation time $M/\gamma$ of the Brownian particle of mass $M$). It ignores the correlations between the particle positions that are present at shorter times and fails for stationary random processes considered in the present paper for all times. In Ref. [20], taking into account the stationarity, it has been obtained from Eq. (2)

$$S(t) = \exp\left[-\gamma_n^2 g^2 \int_0^t \langle x(t')x(0) \rangle (t-t') dt'\right]. \qquad (3)$$

After substituting here $\langle x(t)x(0) \rangle \approx 2Dt$ as the PAF [20], the classical "textbook" expression for the diffusion suppression function was obtained,

$$S(t) = \exp\left[-\frac{1}{3}\gamma_n^2 g^2 D t^3\right]. \qquad (4)$$

Again, the result (4) is correct, but the way to obtain is not. It was used in [20] that $x(t) \approx x(0)$ when, however, the MSD $X(t)$ is zero instead of $2Dt$. Equation (4) was obtained from the incorrect formula (3) for $S(t)$ by using the incorrect expression for the PAF. The incorrectness of Eq. (3), aimed in [20] to describe the random motion of spins for all times $t$ was commented on in Ref. [37]. It can be seen also from the following consideration. Let the free spin-bearing particles exhibit normal diffusion at long times. If they are trapped in a harmonic well with elastic constant $k$, at the times much smaller than the characteristic time $\gamma/k$ the motion of the particles should not be affected by the trap so there is no reason that the influence of the trap would be reflected in $S(t)$. However, the MSD at $t \to 0$ behaves as $X(t) \approx k_B T t^2 / M$ and the PAF for such bounded particles is known to be $\langle x(t) x(0) \rangle = k_B T / k - X(t)/2$ [29, 31]. Thus, in the short-time limit Eq. (3) becomes $S(t) \approx \exp\left[-\gamma_n^2 g^2 \langle x^2 \rangle t^2 / 2\right]$ and depends on $k$ since $\langle x^2 \rangle \approx k_B T / k$. At long times, if $k \to 0$, the MSD becomes $X(t) \approx 2k_B T t / \gamma$. The substitution of $\langle x(t) x(0) \rangle$ in (3) then gives $S(t) \approx \exp\left[-\gamma_n^2 g^2 k_B T t^2 (1/2k - t/6\gamma)\right]$, whereas in the diffusion regime $S(t)$ should be determined by Eq. (4). For unbounded random motion the PAF cannot be used at all because in this case it is ill defined together with the quantity $\langle x^2 \rangle$ [38]. This holds for both the standard (memoryless) Langevin equation [28] describing the Brownian motion and for its generalizations that take into account the effects of memory [32, 33]. The latter case will be considered in Sections 4 and 5, where the GLE is applied to the description of the random motion of spins. Additional arguments will be given against the calculation of the attenuation function by using the PAF. In Section 6, limiting expressions for the NMR signal are given in the case when at long times spin ensembles with memory display anomalous diffusion.

### 3. Revisited formula for the attenuation function

A quantity that should be used in the description of the influence of stochastic motion of spins on the NMR experiments is the well-defined and measurable MSD. The normal diffusion MSD at $t \to \infty$ tends to infinity as $2Dt$, which is not consistent with the approximation $\langle x(t) x(0) \rangle \approx \langle x(t) x(t) \rangle = 2Dt$ used in [20]. When the diffusion is anomalous, $X(t) = Ct^\alpha$, where $C$ is a temperature-dependent parameter, $\alpha = 1$ corresponds to normal diffusion with $C = 2D$, $\alpha < 1$ to sub-diffusion, and $\alpha > 1$ to super-diffusion [39, 40]. To capture all these cases and to obtain the attenuation function that would also be applicable to shorter times in the Brownian motion of spin-bearing particles, Eq. (3) should be modified as follows. The phase accumulation in the rotating frame in Eq. (1) must be calculated through the change of the phase during the time $t$, instead of the phase given by the spin position at time $t$, i.e., through the quantity $\Delta\omega(t) = \gamma_n g \left[x(t) - x(0)\right]$. This follows from the Larmor condition $\omega = -\gamma_n B$ relating the precessional frequency to the applied magnetic field

$B = B_0 + g(t)x(t)$. For the random variable $x(t)$ we do not assume that $x(0) = 0$. Within the concept of accumulating phases [15], the deviation of the precessional phase of a spin at time $t$ from its phase at time $t = 0$ should be thus calculated as

$$\phi(t) = -\gamma_n \int_0^t \left[ B(x(t')) - B(x(0)) \right] dt' = -\gamma_n \int_0^t g \left[ x(t') - x(0) \right] dt'. \tag{5}$$

The difference from the approach used in [17] is that here the actual value of the magnetic field at time $t = 0$ is used, instead of its mean value $\langle B \rangle$, for which there is no reason. By using $\langle B \rangle$, $x(0)$ disappears from the above formula (if $x$ is a random process with $\langle x \rangle = 0$). As will be shown later, this modification of the approach [17] leads to very different results for the attenuation of the transverse magnetization $S = \langle \cos \phi \rangle$. While the formulas for the NMR signal obtained in [16, 17] and the subsequent papers (e.g., [18, 3]) are appropriate only in describing the diffusion of spins (i.e., their motion in the long-time approximation), our approach is equally suitable for the calculation of $S(t)$ at any times. The approximation in [16, 17] can be used to describe the diffusion regime of the particle motion but even for the stationary Markovian (memoryless) Brownian motion described by the standard Langevin theory it must be generalized. The standard Langevin equation can lead to notable corrections to the attenuation of the NMR signal due to diffusion if the frictional time of the Brownian particles is not much smaller than the characteristic time of the experiment (such as the time interval of the spin echo). The effect of memory in the particle dynamics can reveal itself if the memory function present in the GLE does not decay too fast (i.e., when the characteristic time of its decay is comparable to the time of experiment).

For the Gaussian random processes (or small $\phi$) we use Eq. (2), where now

$$\begin{aligned} \langle \phi^2(t) \rangle &= \int_0^t \int_0^t dt' dt'' \langle \omega(t') \omega(t'') \rangle \\ &= \frac{1}{2} \gamma_n^2 g^2 \int_0^t \int_0^t dt' dt'' \left[ X(t') + X(t'') - X(t'' - t') \right]. \end{aligned} \tag{6}$$

Since for stationary processes $X(t)$ is a symmetric function, one can use the following transformation:

$$\int_0^t \int_0^t dt' dt'' X(t'' - t') = 2 \int_0^t dt' (t - t') X(t'). \tag{7}$$

Equations (7), (6) and (2) then give the final simple result [37]

$$S(t) = \exp \left[ -\frac{1}{2} \gamma_n^2 g^2 \int_0^t t' X(t') dt' \right]. \tag{8}$$

This new result is model-independent, applicable for any times and a character of the stochastic motion of spins, providing only that it is stationary and $\phi$ is Gaussian (or small).

Most often, normal diffusion is observed in liquids and gases. Substituting $X(t) \approx 2Dt$ in (8), we return to the classical formula (4). The measured spectral line broadening due to diffusion (half width at half maximum) is $\omega_{1/2} \approx \sqrt{6} a^{1/3}$, where $a = \gamma_n^2 g^2 D / 3$ [1]. At short times the motion of particles is ballistic [31], $X(t) \approx k_B T t^2 / M$, so that Eq. (8) gives

$$S(t) \approx \exp\left[-\frac{k_B T \gamma_n^2 g^2}{8M} t^4\right], \tag{9}$$

and $\omega_{1/2}^2 \approx 4\Gamma(5/4)\Gamma^{-1}(3/4)\left(k_B T \gamma_n^2 g^2 / 8M\right)^{1/2}$, where $\Gamma$ is the gamma function.

### 4. Stochastic motion described by the generalized Langevin equation

As already discussed, the attenuation of the NMR signal within the model of standard Langevin equation [28] for the Brownian motion has been correctly calculated in [25–27]. Here we will describe a more general case, which contains the Langevin theory as a special case. Recently [20], it has been proposed to describe the stochastic motion of spins in gases during the above considered NMR experiment using the GLE, in which the friction force is modeled by the convolution of the exponentially decaying memory kernel [34, 35, 40] $\Gamma(t) = (\gamma^2 / m)\exp(-\gamma t / m)$ with the particle velocity $\upsilon(t) = \dot{x}(t)$,

$$M\dot{\upsilon} + \int_0^t \Gamma(t - t')\upsilon(t')dt' = f(t). \tag{10}$$

Here, $f(t)$ is a stochastic force, $m \ll M$ is the mass of molecules in the surrounding medium and $\gamma$ is the friction coefficient proportional to the medium viscosity. By the fluctuation-dissipation theorem, the stochastic force describes colored noise, $\langle f(0)f(t)\rangle = k_B T \Gamma(t)$ [33]. Equation (10) is easily solved by the method presented in [40, 41] and used for a similar problem in [35]. The solution for the velocity autocorrelation function $\langle \upsilon(t)\upsilon(0)\rangle$ determines the time-dependent diffusion coefficient $v(t) = \int_0^t \langle \upsilon(0)\upsilon(\tau)\rangle d\tau$,

$$v(t) = \frac{k_B T}{M}\left\{\frac{\gamma}{m\zeta_-\zeta_+} - \frac{1}{\zeta_+ - \zeta_-}\left[\left(1 - \frac{\gamma}{m\zeta_+}\right)e^{-\zeta_+ t} - \left(1 - \frac{\gamma}{m\zeta_-}\right)e^{-\zeta_- t}\right]\right\}, \tag{11}$$

where $\zeta_{-,+} = (\gamma/2m)\left(1 \mp \sqrt{1 - 4m/M}\right)$. Note that in Ref. [20] the second term in {} was incorrectly obtained with the opposite sign. (It is easily seen, e.g., by taking the limit $m/M \to 0$, when the memory integral in (10) is replaced by $\gamma\upsilon(t)$ and one must get the result from the standard Langevin theory of the Brownian motion [28]. In this case $v(t)$ is

$$v(t) = \frac{k_B T}{\gamma}\left[1 - \exp\left(-\frac{\gamma}{M}t\right)\right], \tag{12}$$

while, by using $\zeta_+ \to \gamma/m$, $\zeta_- \to \gamma/M$, $\zeta_+ - \zeta_- \to \gamma/m$ (if $m/M \to 0$), and $\zeta_+\zeta_- = \gamma^2(mM)^{-1}$ (for all $m$ and $M$), the result from [20] gives the sign + before the second term in the brackets. It is claimed in [20] that from $v(t)$ the PAF has been found having the form

$$\xi(t) = \frac{k_B T}{M(\zeta_+ - \zeta_-)}\left[\frac{1}{\zeta_+}\left(1 - \frac{\gamma}{m\zeta_+}\right)e^{-\zeta_+ t} - \frac{1}{\zeta_-}\left(1 - \frac{\gamma}{m\zeta_-}\right)e^{-\zeta_- t}\right]. \tag{13}$$

Again, this is not correct for the following reasons. By integrating $2v(t)$ [35, 40, 41], one obtains the MSD,

$$X(t) = \frac{2k_B T}{\gamma}t - \frac{2k_B T}{\gamma^2}(M - m) + 2\xi(t). \tag{14}$$

It has been used that $\zeta_+ + \zeta_- = \gamma/m$ and $\zeta_+\zeta_- = \gamma^2(mM)^{-1}$. Equation (14) corrects also the solution of the GLE that has been obtained earlier in [25] and then used to calculate the attenuation of the NMR spin echo.

At $t \to \infty$ and $t \to 0$, respectively, one comes from (14) to the already used formulas $X(t) \approx 2k_B T t/\gamma$ and $X(t) \approx k_B T t^2/M$. The function $\xi(t)$ cannot be identified with $\langle x(t)x(0)\rangle$. Let us assume that $\xi(t) = \langle x(t)x(0)\rangle$. Simultaneously we have $X(t) = 2\langle x^2\rangle - 2\langle x(t)x(0)\rangle$, so that Eq. (14) can be rewritten to $X(t) = k_B T \gamma^{-1} t - k_B T \gamma^{-2}(M - m) + \langle x^2\rangle$. This relation does not correspond to the correct solution for $X(t)$, as it is seen already from the MSD long and short time limits or from the fact that $\langle x^2\rangle$ should be a constant. An evidently wrong expression $S(t) \propto \exp(\gamma_n^2 g^2 \kappa t)$ with $\kappa = k_B T M^2 \gamma^{-3}$ follows from (13) and (3) at $m \ll M$ and $\gamma t/m \gg 1$. If $\gamma t/m \ll 1$, $\xi(t)$ in the main approximation is $\xi(t) \approx k_B T M \gamma^{-2}$, so that at short times we have a decay of $S(t)$, as it should be, but at long times we have not, which again shows incorrectness of $S(t)$ found in [20]. By substituting (14) in (8), the new formula for $S(t)$ at long times reads

$$S(t) \approx \exp\left\{-\frac{k_B T \gamma_n^2 g^2}{3\gamma}\left[t^3 - \frac{3}{2\gamma}(M - m)t^2 + \frac{3M^3}{\gamma^3}\left(1 - \frac{3m}{M} + \frac{m^2}{M^2}\right)\right]\right\}. \tag{15}$$

This equation determines corrections to the classical result (4), providing the used GLE model is applicable. The linewidth broadening corresponding to (15) is given mainly by the law $\omega_{1/2} \sim \left(\gamma_n^2 g^2 k_B T / 3\gamma\right)^{1/3}$. If necessary, corrections to $\omega_{1/2}$ can be determined from (15) for a concrete system. Having a model for the friction coefficient $\gamma$ and the viscosity $\eta$, the temperature dependence of $\omega_{1/2}$ can be predicted. For example, in gases at high temperatures $\eta \sim \sqrt{T}$ [42]. Assuming the validity of the Stokes formula for $\gamma$, $\gamma = 6\pi\eta R$ ($R$ is the particle radius), this gives $\omega_{1/2} \sim T^{1/6}$ (with a correction that decreases with $T$ as $\sim T^{-1/6}$) instead of $\sim T^{-1/2}$ found in [20]. At low temperatures $\eta \sim T^{3/2}$ and $\omega_{1/2} \sim T^{-1/6}$. Note however that gases are systems where the memory in the particle dynamics is of low significance and the Brownian motion is very well described by the standard Langevin equation [30]. The considered GLE model seems to be applicable to viscoelastic fluids [34].

Equation (14) at $m \ll M$ becomes the MSD within the standard Langevin theory [28]. With this solution, Eq. (8) gives exactly the result obtained by Stepišnik [25], which at long times agrees with (15). In the case $4m > M$ the roots $\xi_{+,-}$ are complex and the solution (14) describes damped oscillations [35]. In the overdamped limit $4m \gg M$ one obtains from (14)

$$X(t) \approx 2D\left\{t + \frac{m}{\gamma}\left[1 - \exp\left(-\frac{\gamma t}{2m}\right)\cos\left(\frac{\gamma t}{\sqrt{mM}}\right)\right]\right\}. \tag{16}$$

This equation significantly differs from Eq. (30) in Ref. [25] obtained in the limit of large correlation times of the particles of mass $m$, surrounding the particles of mass $M$.

## 5. Hahn spin echo

Modern NMR pulse sequences come from the simple refocusing principle of the spin echo developed by Hahn [43]. In this experiment, at time $t = \tau$ after the first 90° rf pulse at $t = 0$ the spin phases are inverted by a 180° pulse. Measurements of the echo signal amplitude at time $2\tau$ allow accurate determining of the diffusion coefficients of nuclear spins. During the experiment, a static magnetic field that creates macroscopic magnetization along the axis $x$ and a constant magnetic field gradient $g$ are applied. As in Section 3, we expressed the attenuation of the signal due to the stochastic motion of spins (2) through the accumulation of the changes of spin phases $\phi(t)$. Now we have

$$\langle \phi^2(t) \rangle = \gamma_n^2 g^2 \left\langle \left\{ \int_0^\tau [x(t') - x(0)]dt' - \int_\tau^t [x(t') - x(0)]dt' \right\}^2 \right\rangle. \tag{17}$$

The sign before the second integral accounts for the fact that at time $\tau$ all phases are inverted. Equation (17) can be again expressed through the MSD. After the averaging and use of the stationary condition one finds

$$\langle \phi^2(t) \rangle = \gamma_n^2 g^2 \int_0^t dt'(t' - 2\tau) X(t') + 2\int_0^\tau dt'(2t' - t) X(t') + 2\int_0^t dt' \int_0^\tau dt'' X(t' - t''). \tag{18}$$

Other equivalent forms of Eq. (18) are possible as well. In the special case of the Einstein diffusion we get from Eqs. (18) and (2) at $t = 2\tau$ the famous Stejskal-Tanner formula [44]

$$S(2\tau) = \exp\left[-\frac{2}{3}\gamma_n^2 g^2 D\tau^3\right]. \tag{19}$$

At arbitrary $t > \tau$

$$S(t) = \exp\left[-\frac{1}{3}\gamma_n^2 g^2 D(t^3 - 6t\tau^2 + 6\tau^3)\right]. \tag{20}$$

It is interesting that the maximum of the function $S(t)$ is not at the echo time $2\tau$ but earlier, at $t = \sqrt{2}\tau$. This has been for the first time obtained and experimentally verified in [22, 23].

Often the pulsed gradient method is used. Let the first gradient pulse begins at time $t = \tau_g$ after the 90° rf pulse and the second one at time $t = \Delta$ [10]. The 180° rf pulse is applied between these gradient pulses, the duration of each is $\delta$. The result for $\tau_g + \delta < \tau$ (up to the second rf pulse) is the same as for a steady gradient (Eq. (8) with $t = \delta$). Due to stationarity, after the second rf and gradient pulses the result of calculations also does not depend on $\tau_g$ (as distinct from [45]) and can be evaluated from

$$S(\delta, \Delta) = \exp\left\{-\frac{1}{2}\gamma_n^2 g^2 \left[\int_0^\delta dt' \int_0^\delta dt'' X(t'' - t' + \Delta) - 2\int_0^\delta dt'(\delta - t') X(t')\right]\right\}. \tag{21}$$

This new formula simplifies to the well-known relation [3] when the MSD is $X(t) = 2Dt$,

$$S(\delta, \Delta) = \exp\left[-\gamma_n^2 g^2 D \delta^2 (\Delta - \delta/3)\right]. \tag{22}$$

As it will be shown below, for short-time pulses with $\delta \ll \Delta$ this expression can be used also to describe the signal from anomalously diffusing particles [17, 3]. Generally, however, this is not true. If $\delta = \Delta = \tau$ is substituted in (21), we obtain the damping of the signal at the echo time $2\tau$, in the case of the steady gradient, $S(\tau, \tau) = S(2\tau)$ which agrees with equations (2) and (18). The new exact expression for $S(t)$ within the GLE model described in the preceding section is obtained by substituting the MSD (14) in Eq. (21):

$$\frac{-\ln S(\delta, \Delta)}{(\gamma_n g)^2 D} = \delta^2\left(\Delta - \frac{\delta}{3}\right) + \frac{\gamma}{M}\frac{1}{\xi_+ - \xi_-}\left[\varphi(\xi_+) - \varphi(\xi_-)\right], \tag{23}$$

where

$$\varphi(z) = \frac{1}{z}\left(1 - \frac{\gamma}{mz}\right)\left\{-\frac{2\delta}{z} + \frac{2}{z^2}\left[1 - e^{-z\Delta} - e^{-z\delta} + \frac{1}{2}\left(e^{-z(\Delta+\delta)} + e^{-z(\Delta-\delta)}\right)\right]\right\}. \tag{24}$$

At long times Eq. (23) differs from (22) obtained for normal diffusion. The difference depends on the relation between the frictional time $M/\gamma$ and the experimental times $\delta$ and $\Delta$. In the main approximation for $\delta \gg m/\gamma$ and by using $m/M \to 0$, the result (23) converts to

$$\frac{-\ln S(\delta,\Delta)}{(\gamma_n g)^2 D} \approx \delta^2\left(\Delta - \frac{\delta}{3}\right) - 2\delta\left(\frac{M}{\gamma}\right)^2 + 2\left(\frac{M}{\gamma}\right)^3. \tag{25}$$

Formally this is the same result as in the model described by the traditional Langevin equation for particles of mass $M$ [25]. However, while in the GLE model the correction to the attenuation function for normal diffusion can be notable, the long-time approximation in the Langevin model without memory requires $\delta \gg M/\gamma$, when the second and third terms in the right-hand side of Eq. (25) are very small. In its general form Eq. (23) significantly differs from the result in [25] for $S(t)$ obtained from the solution of the GLE (10), which was corrected in the preceding section.

## 6. Anomalous diffusion

The first attempt to describe the NMR experiments on systems displaying anomalous diffusion has been published by Jug [45]. Soon it has been shown [16, 17] that the results of [45] contradict the principle of time invariance and the correct expressions for the NMR spin echo attenuation $S(t)$ due to diffusion have been obtained. Assuming the MSD of the form (see Sec. 3) $X(t) = Ct^\alpha$, and by using (8), it is easy to find $\langle \phi^2(t) \rangle$ and then $S(t)$ from Eq. (2) that generalizes the induction signal (4):

$$S(t) = \exp\left(-\frac{1}{2}\frac{\gamma_n^2 g^2 C}{\alpha+2} t^{\alpha+2}\right). \tag{26}$$

$S(t)$ in the Hahn echo experiment is readily obtained from (18) if, for symmetrical $X(t)$, the last integral is rewritten as

$$\int_0^\tau dt' \int_{-t'}^{t-t'} dx\,|x|^\alpha = \int_0^\tau dt'\left[\int_{-t'}^0 dx(-x)^\alpha + \int_0^{t-t'} x^\alpha dx\right] = \int_0^\tau dt'\left[\int_0^{t'} x^\alpha dx + \int_0^{t-t'} x^\alpha dx\right].$$

The final result of integration in (18) is

$$\frac{\langle \phi^2(t)\rangle}{C\gamma_n^2 g^2} = \frac{1}{(\alpha+1)(\alpha+2)}\left[t^{\alpha+2} - 2\tau^{\alpha+2} + (\alpha+2)(t-2\tau)(t^{\alpha+1} - 2\tau^{\alpha+1}) - 2(t-\tau)^{\alpha+2}\right]. \tag{27}$$

The minimum of $\langle \phi^2(t)\rangle$ is not at $t = 2\tau$ but earlier. It can be found from the equation

$$(\alpha+3) y^{\alpha+1} - 2(y-1)^{\alpha+1} - 2(\alpha+1) y^\alpha - 2 = 0, \quad y \equiv t/\tau > 1.$$

Only for $\alpha = 0$ one finds $y = 2$ (the echo time). For $\alpha = 1$ (normal diffusion) and 2 (ballistic motion), $y = \sqrt{2}$. For $\alpha \in (0, 2)$, $y$ changes from 2 to about $\sqrt{2}$ (with increasing $\alpha$ the minimum time decreases to $\sqrt{2}$ at $\alpha = 1$, then slightly increases and again decreases to $\sqrt{2}$ at $\alpha = 2$). From Eqs. (2) and (27), a simple formula follows at the spin echo time $2\tau$,

$$S(2\tau) = \exp\left[-2\gamma_n^2 g^2 C \frac{2^\alpha - 1}{(\alpha+1)(\alpha+2)} \tau^{\alpha+2}\right] \tag{28}$$

At $\alpha = 1$ we return to Eq. (19). In the case of pulsed gradient echo, $S(t)$ from (21) is [17]

$$S(\delta, \Delta) = \exp\left\{-\frac{C\gamma_n^2 g^2}{2(\alpha+1)(\alpha+2)}\left[(\Delta+\delta)^{\alpha+2} + (\Delta-\delta)^{\alpha+2} - 2\Delta^{\alpha+2} - 2\delta^{\alpha+2}\right]\right\}. \tag{29}$$

This equation significantly differs from the result

$$S(\delta, \Delta) = \exp\left\{-\frac{1}{2}C\gamma_n^2 g^2 \delta^2\left[(\Delta-\delta)^\alpha + \frac{2\delta^\alpha}{2+\alpha}\right]\right\} \tag{30}$$

obtained in Ref. [27] and shows that the calculation of $\langle \Delta\phi^2(t)\rangle$ for the spin echo in [27], Eq. (43), the details of which are given in Appendix B, is not correct. When $\delta = \Delta = \tau$, from (29) we return to the result for steady gradient. When $\delta \ll \Delta$, for $X(\Delta) = C\Delta^\alpha$ (29) can be given the same form as for normal diffusion, $S(\Delta,\delta) \approx \exp\left[-\gamma_n^2 g^2 \delta^2 X(\Delta)/2\right]$ [17, 3].

## 7. Conclusion

The description of the stochastic motion of particles in liquids based on the memoryless Brownian motion theory is often inappropriate to interpret observations. In the last decade this has been clearly shown mainly in the studies of the motion of microparticles in optical trapping experiments [29–31]. These experiments have proven that the standard Einstein and Langevin theories fail to describe the observations, except long times when the particles are in the diffusion regime. At the same time, theoretical descriptions of various NMR experimental methods, which are for more than half a century successfully applied to study the diffusion and self-diffusion processes, usually do not go beyond the long-time limit and the memoryless description of the dynamics of spin-bearing particles. A few attempts based on the generalized Langevin equation to take into account the memory effects that are revealed at shorter times were not successful [25, 27, 20]. As we have shown in this paper, either this equation was not solved correctly, as in [25], or the calculations of the attenuation function $S(t)$ are mistaken [27]. In the recent paper [20] both the solution of the GLE and the result for $S(t)$ are erroneous. The searched positional autocorrelation function for spin-bearing particles is not a proper quantity to describe the attenuation function of the NMR signal due to the stochastic motion of particles. For unbounded stationary motion this function even cannot be defined. In the present work, the attenuation function was evaluated for two examples of the NMR experiments: when the nuclear induction signal is measured in the presence of a field gradient, and for the steady and pulsed gradient Hahn spin echo. The observed attenuation is calculated through the accumulation of the spin phases in the frame rotating with the resonance frequency. Coming from the changes of the phases during the time of observation, this accumulation is represented through the mean square displacement of spins in stationary and Gaussian random processes. Several new formulas have been obtained that are valid for any times of measurements. At long times they give known results in the case of normal diffusion and describe the influence of the signal by anomalous diffusion, but are equally applicable to the stochastic motion described by other models, e.g., by various generalizations of the Langevin theory of the Brownian motion. We have considered in detail the model based on the generalized Langevin equation with memory exponentially decaying in time.

## Acknowledgement

This work was supported by the Agency for the Structural Funds of the EU within the project NFP 26220120033, and by the Ministry of Education and Science of the Slovak Republic through Grant VEGA No. 1/0348/15.